\documentclass[%        Class options:
aps,%                   American Physical Society
prd,%                   Physical Review D
preprint,%              Preprint layout
tightenlines,%          Single spaced lines
superscriptaddress,%    Authors' addresses linked with superscripts
nofootinbib,%           Does not treat footnotes as references
floatfix,%              Fixes float errors
a4paper]%               A4 paper format
{revtex4}%              REVTEX 4 Package used
\usepackage{graphicx}%  Default Latex 2eps package for embedding figures
\begin{document}
%........................................................................
\begin{flushright}
{\small BARI-TH 452/02}
\end{flushright}
\title{
Gravitational Field of Static Thin Planar Walls \\  in Weak-Field
Approximation}
\author{        L.~Campanelli}
\affiliation{   Dipartimento di Fisica
                and Sezione INFN di Bari\\
                Via Amendola 173, 70126 Bari, Italy\\}
\author{        P.~Cea}
\affiliation{   Dipartimento di Fisica
                and Sezione INFN di Bari\\
                Via Amendola 173, 70126 Bari, Italy\\}
\author{        G.L.~Fogli}
\affiliation{   Dipartimento di Fisica
                and Sezione INFN di Bari\\
                Via Amendola 173, 70126 Bari, Italy\\}
\author{        L.~Tedesco}
\affiliation{   Dipartimento di Fisica
                and Sezione INFN di Bari\\
                Via Amendola 173, 70126 Bari, Italy\\}
\vskip 1truecm
\begin{abstract}
%............................................................
We investigate gravitational properties of  thin planar wall
solutions of the Einstein's equations in the weak field
approximation. We find the general metric solutions and discuss
the behavior of a particle placed initially at rest to one side of
the plane. Moreover we study the case of non-reflection-symmetric
solutions.
%.............................................................
\end{abstract}
\medskip
\maketitle
\section{Introduction}
The study of the gravitational properties of topological defects
from a cosmological point of view is not an academic exercise
\cite {VILENKINBOOK}. Topological defects can be point-like
(monopoles), string-like (cosmic strings) or planar (domain
walls), depending on the particle physics model; also important
defects are textures, point-like  defects in space-time. In
general  topological defects are regions in space in which the
energy density is trapped. These regions of "surplus" of energy
might act as seed for formation of structure, by means of the
gravitational accretion mechanism \cite {BRANDENBERGER1}.
Moreover,  if the seeds for galaxy formation are indeed provided
by topological defects, one  expects that the resulting power
spectrum of cosmic background radiation anisotropy to differ from
that predicted by inflation \cite {TUROK}. Therefore it is
important to investigate these cosmological  objects from a
gravitational point of view. On the other hand many of the
proposed observational tests for the existence of topological
defects are based on their gravitational interactions. The more
relevant aspect is connected with the unusual sources of gravity
of topological defects. For example, the gravitational field
around a static string is very peculiar \cite {VILENKIN:81}. In
fact when we consider the Newtonian limit of Einstein's equations
with a source term given by $T^{\mu}_{\nu} = \text{diag} (\sigma,
-p_1, -p_2, -p_3)$, we find that the Newtonian potential $\Phi$
satisfies  $\nabla^2 \Phi= 4 \pi G (\sigma+p_1+p_2+p_3)$, the
pressure terms also contribute to the "gravitational mass". If the
string is infinite in the $x$ direction, we have $p_1=-\sigma$, in
other terms it has a large relativistic tension (negative
pressure). For $y$ and $z$ directions we have $\nabla^2 \Phi =0$.
All this indicates that space out of the infinite straight cosmic
string is flat and therefore a test particle does not feel any
gravitational attraction. In general cosmic strings do not produce
gravitational force on the surrounding matter but they deflect
light \cite{A, B, C,D,E,F,G}, while domain walls are repulsive but
do not deflect light. In any case the metric describing the
exterior of a gauge cosmological string was derived by Vilenkin
\cite {VILENKIN:81} linearizing Einstein's equations. Later the
solution was shown to be exact and unique in the full non linear
theory of general relativity \cite{{HISCOCK},{LINET}}.
\\
Moreover the gravitational properties of domain walls are
important due to their striking cosmological implications. The
repulsive gravitational properties of domain walls lead to the
interesting proposal that the energy density of the Universe could
be  dominated by a network of low-tension domain walls
\cite{Friedland,Bucher,Battye,Fabris}. This is in relation to the
recent evidence  \cite{Perlmutter,Riess}  of an accelerating
Universe from the distance measurements of supernovae  Ia with
redshift between 0 and 1. This proposed is alternative to the two
most famous candidates of dark energy: cosmological constant and
quintessence.
\\
The metric for infinite, static plane-symmetric domain wall is
rather peculiar and is different from that of a massive plane. The
energy-momentum tensor of a wall lying in the (y-z)-plane is
\begin{equation}
\label{eq01}
T^{\mu}_{\, \nu} = \sigma \delta(x) \text{diag} (1,0,1,1),
\end{equation}
where $\sigma$ is the energy density of the wall.
Vilenkin \cite {VILENKIN:81} studied the gravitational
 properties
of static domain walls and strings in the linear approximation of
general relativity. The Newtonian limit of the Einstein equations
for domain walls is $\nabla^2 \phi= -4 \pi G \sigma$, that is to
say the domain walls are repulsive. The only static solution with
$T^{\mu}_{\, \nu}\neq 0$ in  $x=0$ and $T^{\mu}_{\, \nu}= 0$
everywhere, is \cite {TAUB}
\begin{equation}
\label{eq02}
ds^2=\frac {1}  {\sqrt{1- k|x|}} \,  (dt^2-dx^2) -
(1-k|x|)(dy^2-dz^2)
\end{equation}
with $T^{\mu}_{\nu} = \frac {k} {8 \pi G} \delta(x) \text {diag}
(1,0,1/4,1/4)$. Note that this form of $T^{\mu}_{\nu}$ is not
consistent with Eq.~(\ref{eq01}). Therefore is not possible to
have a static gravitational field for a planar wall. Indeed, the
time-dependent solution of Einstein's equation in the case of
planar domain walls was found by Vilenkin \cite {VILENKIN:83} and
by Ipser and Sikivie \cite{E}. They found the following metric:
\begin{equation}
\label{eq03} ds^2= (1- 2 \pi G \sigma |x|)^2 dt^2 - dx^2 - (1- 2
\pi G \sigma |x|)^2 e^{4 \pi G \sigma t} (dy^2+dz^2),
\end{equation}
obtained in the thin wall approximation. The gravitational field
of thick walls have been studied by Widrow \cite {WIDROW}.
\\
In this paper we investigate the gravitational properties if thin
planar domain walls by solving Einstein's equations in the weak
field approximation. In Section II we discuss the general solution
of linearized Einstein's equations for static domain walls,
ferromagnetic domain walls and shell dust. Section III is devoted
to non-reflection-symmetric solutions. Finally in Section IV we
draw our conclusions.
%
%##############################################################################################
%
%
%
\vskip 2truecm
\section{Gravitational field of  thin planar wall in
 weak-field
\\
approximation}
\vskip 1truecm
In previous papers the study of gravitational field of planar wall
has  been made in the framework of thin-wall approximation with
planar symmetry. The thin wall limit permits, in four dimensional
space-time, to treat the wall as an infinitely thin (2+1)-surface.
The weak field approximation gives an interesting feature of the
wall's gravitational field. In fact it does not correspond to any
exact static solution of the Einstein's equation. The
reflection-symmetric solutions by Vilenkin are inconsistent with
the Kasner  solution \cite{LANDAU}.
\\
Let us write the metric tensor as a flat-space Lorentz metric
$\eta_{\mu \nu}=\text{diag} (1,-1,-1,-1)$ plus a perturbation term
$h_{\mu \nu}$:
\begin{equation}
\label{eq1} g_{\mu \, \nu} = \eta_{\mu \, \nu} + h_{\mu \, \nu} ,
\end{equation}
where $|h_{\mu \nu}|\ll 1$ everywhere in space-time. In this
approximation the linearized Einstein's field equations read
\begin{equation}
\label{eq2}
\left( \nabla^{2} - \partial_{t}^{\;2} \right) h_{\mu
\, \nu} = 16\pi G \! \left( T_{\mu \, \nu} -
\mbox{$\frac{1}{2}$}\, \eta_{\mu \, \nu} T \right)\!,
\end{equation}
where $T=T^{\mu}_{\mu}$ denotes the trace of the energy equations
of the linearized theory and $h^{\mu}_{\mu}$ satisfies the so
called harmonic gauge or de Donder gauge conditions:
\begin{equation}
\label{eq3}
\partial_{\mu} \! \left( h_{\;\; \nu}^{\,\mu} -
\mbox{$\frac{1}{2}$} \, \delta_{\;\; \nu}^{\,\mu} \, h \right)  =
0.
\end{equation}
The remaining coordinates are restricted to the transformation:
\begin{equation}
\label{eq4}
h_{\mu \nu}' = h_{\mu \nu} - \xi_{\mu,\,\nu} - \xi_{\nu \!,\:\mu},
\end{equation}
with
\begin{equation}
\label{eq5}
(\partial_{t}^2 - \nabla^2)\xi_{\nu} = 0.
\end{equation}
In the limit of vanishing thickness the planar configuration  can
be considered as an infinitely thin three-dimensional hypersurface
with the energy-momentum tensor $T^{\mu}_{\nu}$ having in general
singularities on  the hypersurface.  The energy-momentum tensor
for thin homogeneous massive plane is taken to be
\begin{equation}
\label{eq6}
T_{ \; \nu}^ {\mu} = \delta(x) \; \text{diag} \, (\sigma,0,-p,-p),
\end{equation}
where $\sigma$ is the energy density of the wall and $p$ the
pressure density; the plane is at the surface $x=0$, the $x$-axis
being perpendicular to the wall.
\\
Eqs. (\ref{eq2}) and (\ref{eq3}) with the tensor (\ref{eq6}),
give
\begin{eqnarray}
\label{eq7}
h_{00} \!\! & = & \!\! A t \, + \, 4\pi G \, (\sigma + 2p) |x|,
\\
\label{eq8}
h_{11} \!\! & = & \!\! B t \, + \, 4\pi G \, (\sigma - 2p) |x|,
\\
\label{eq9}
h_{22} \!\! & = & \!\! h_{33} = C t \, + \, 4\pi G \, \sigma |x|,
\end{eqnarray}
where $A$, $B$ and $C$ are constants and $h_{\mu \nu}=0$ for $\mu
\neq \nu$. If we impose the harmonic gauge conditions (\ref
{eq3}), we have $C=- \frac {1} {2} (A+B)$, therefore the  non
static massive plane metric in the linear approximation is:
\begin{eqnarray}
\label{eq10}  ds^2 & = & [1+A t+4\pi G(\sigma + 2 p)|x|]dt^2+ [1+B
t+4 \pi G (\sigma - 2 p)|x|] dx^2 + \nonumber
\\
 & + & \left[- \frac {A+B} {2} + 4 \pi G \sigma |x|\right]
(dx^2+dz^2).
\end{eqnarray}
which depends on two constants $A$ and $B$.
\\
Taub \cite{TAUB} has shown that, in the vacuum, any
plane-symmetric metric can be written as:
\begin{equation}
\label{eq11}
ds^2 = e^{2u} (dt^2 -dx^2) \, - \, e^{2v}(dy^2 + dz^2),
\end{equation}
where $u$ and $v$ are functions of $t$ and $x$, and from symmetry
of the problem we have $u(-x,t)=u(x,t)$ and $v(-x,t)=v(x,t)$. The
functions $u$ and $v$ satisfy the following conditions:
\begin{eqnarray}
\label{eq12}
e^{2v} \!\!&=&\!\! f(x+t)\,+\, g(x-t), \\
\label{eq13}
u \!\!&=&\!\! - \mbox{$\frac{1}{2}$}\,v \,+\, h(x+t)\,+\, k(x-t),
\end{eqnarray}
where $f$ and $g$ are arbitrary functions of the variables
indicated, and the functions $h$ and $k$ satisfy the following
conditions:
\begin{eqnarray}
\label{eq14}
h(x+t) \!\!&=&\!\!
    \left\{ \begin{array}{ll}
           \mbox{$\frac{1}{2}$} \ln{f'} + \mbox{constant} & \mbox{if $f' \neq 0$}, \\
           \mbox{arbitrary} & \mbox{if $f' = 0$},
    \end{array}
    \right. \\
\label{eq15}
 k(x-t) \!\!&=&\!\!
    \left\{ \begin{array}{ll}
           \mbox{$\frac{1}{2}$} \ln{g'} + \mbox{constant} & \mbox{if $g' \neq 0$}, \\
           \mbox{arbitrary} & \mbox{if $g' = 0$},
    \end{array}
    \right.
\end{eqnarray}
where primes indicate derivatives of the functions with respect to
its arguments.
\\
In the weak-field approximation, that is $|u| \ll 1$ and $|v| \ll
1$, from Eq. (\ref{eq11}) we have:
\begin{eqnarray}
\label{eq16}
h_{00} \!\!&=&\!\! 2u, \\
\label{eq17}
h_{11} \!\!&=&\!\! -2u, \\
\label{eq18}
h_{22} \!\!&=&\!\! h_{33} = - 2v,
\end{eqnarray}
all others $h_{\mu \, \nu}$ being  zero. These $h_{\mu \, \nu}$
cannot be directly compared with Eqs.
(\ref{eq7})--(\ref{eq8})--(\ref{eq9}) since they satisfy the
harmonic coordinate conditions (\ref{eq3}) only if $v$ is
constant. But, by comparing Eq. (\ref{eq9}) with Eq. (\ref{eq18})
we see this it is not the case.
\\
A coordinate transformation (\ref{eq4}) with
\begin{eqnarray}
\label{eq19}
\xi_{0} \!\!&=&\!\! \alpha \, t^2, \\
\label{eq20}
\xi_{1} \!\!&=&\!\! \beta \, x^2 \,\mbox{sign}x, \\
\label{eq21}
\xi_{2} \!\!&=&\!\! \xi_{3} = 0,
\end{eqnarray}
where $\alpha$ and $\beta$ are free parameters, brings the metric
(\ref{eq16}),(\ref{eq17}) and (\ref{eq18}) to the form:
\begin{eqnarray}
\label{eq22}
h_{00}' \!\!&=&\!\! 2u - 4\alpha \, t, \\
\label{eq23}
h_{11}' \!\!&=&\!\! -2u -4\beta \, |x|, \\
\label{eq24}
h_{22}' \!\!&=&\!\! h_{33}' = - 2v.
\end{eqnarray}
These $h_{\mu \, \nu}'$ satisfy the harmonic coordinate conditions
(\ref{eq3}) if and only if $\partial_{t}v = -\alpha$ and $
\partial_{x}v = \beta \,\mbox{sign}x$, that is
\begin{equation}
\label{eq25}
v = \beta |x| -\alpha t.
\end{equation}
Thus, the coordinate transformation (\ref{eq19}),(\ref{eq20}) and
(\ref{eq21}) brings the metric (\ref{eq16}), (\ref{eq17}) and
(\ref{eq18}) to an harmonic form:
\begin{eqnarray}
\label{eq26}
h_{00}' \!\!&=&\!\! 2u - 4\alpha \, t, \\
\label{eq27}
h_{11}' \!\!&=&\!\! -2u - 4\beta \, |x|, \\
\label{eq28} h_{22}' \!\!&=&\!\! h_{33}' = - 2\beta \, |x| +
2\alpha \, t.
\end{eqnarray}
In the weak-field approximation, Eq. (\ref{eq12}) became
\begin{equation}
\label{eq29}
1+2v = f(x+t)\,+ \, g(x-t).
\end{equation}
Now, for simplicity, we restrict to the half-space $x>0$. Since
the functions $u$ and $v$ are linear in the variable $x$ and $t$,
we can write:
\begin{eqnarray}
\label{eq30}
f(x+t) \!\!&=&\!\! a_{1}(x+t) + 1, \\
\label{eq31}
g(x-t) \!\!&=&\!\! a_{2}(x-t),
\end{eqnarray}
where $a_{1}$ and $a_{2}$ are constants. Inserting Eqs.
(\ref{eq30}) and (\ref{eq31}) in Eq. (\ref{eq29}) we have:
\begin{equation}
\label{eq32}
2v = (a_{1}+a_{2})x + (a_{1}-a_{2})t.
\end{equation}
Now, there are three cases, that is:
\begin{eqnarray}
\label{eq33}
a_{1} \!\!& \neq &\!\! 0 \;,\;\;\; a_{2} \neq 0, \\
\label{eq34}
a_{1} \!\!& = &\!\! 0 \;,\;\;\; a_{2} \neq 0, \\
\label{eq35}
a_{1} \!\!& \neq &\!\! 0 \;,\;\;\; a_{2} = 0.
\end{eqnarray}
By using Eqs. (\ref{eq12})-(\ref{eq15}), Eq. (\ref{eq25}) and Eq.
(\ref{eq32}), we find for the three case, respectively:
\begin{equation}
\label{eq36}
u = \mbox{$\frac{1}{2}$} \,v,
\end{equation}
\begin{equation}
\label{eq37}
\left\{ \begin{array}{ll}
           u = \mbox{$\frac{1}{2}$} \,v + c_{1}(x+t), \\
          \alpha = \beta,
    \end{array}
    \right.
\end{equation}
and
\begin{equation}
\label{eq38}
\left\{ \begin{array}{ll}
           u = \mbox{$\frac{1}{2}$} \,v + c_{2}(x-t), \\
          \alpha = -\beta,
    \end{array}
    \right.
\end{equation}
where $c_{1}$ and $c_{2}$ are constants. Inserting Eqs.
(\ref{eq36}), (\ref{eq37}) and (\ref{eq38}) in Eqs.
(\ref{eq26})-(\ref{eq28}), we have respectively:
\begin{eqnarray}
\label{eq39}
h_{00}' \!\!&=&\!\! - 5\alpha \, t + \beta \,x, \\
\label{eq40}
h_{11}' \!\!&=&\!\! -5\beta \, x + \alpha \, t, \\
\label{eq41}
h_{22}' \!\!&=&\!\! h_{33}' = 2\alpha \, t - 2\beta \, x,
\end{eqnarray}
\begin{eqnarray}
\label{eq42}
h_{00}' \!\!&=&\!\! (2c_{1} - 5\alpha) \, t + (2c_{1} + \alpha) \,x, \\
\label{eq43}
h_{11}' \!\!&=&\!\! (\alpha - 2c_{1}) \, t - (2c_{1} + 5\alpha) \,
x, \\
\label{eq44}
h_{22}' \!\!&=&\!\! h_{33}' = 2\alpha \, t - 2\alpha \, x,
\end{eqnarray}
and
\begin{eqnarray}
\label{eq45}
h_{00}' \!\!&=&\!\! -(2c_{2} + 5\alpha) \, t + (2c_{2} - \alpha) \,x, \\
\label{eq46}
h_{11}' \!\!&=&\!\! (2c_{2} + \alpha) \, t + (5\alpha - 2c_{2}) \,
x, \\
\label{eq47}
h_{22}' \!\!&=&\!\! h_{33}' = 2\alpha \, t + 2\alpha \, x.
\end{eqnarray}
Comparing the metrics (\ref{eq42})-(\ref{eq44}) and
(\ref{eq45})-(\ref{eq47}) with the metric (\ref{eq7})-(\ref{eq9}),
we find respectively:
\begin{eqnarray}
\label{eq48}
A \!\!&=&\!\! 8\pi G\, (2\sigma+p), \\
\label{eq49}
B \!\!&=&\!\! -8\pi G\, (\sigma+p), \\
\label{eq50}
\alpha \!\!&=&\!\! -2\pi G\, \sigma, \\
\label{eq51}
c_{1} \!\!&=&\!\! \pi G\, (3\sigma+4p),
\end{eqnarray}
and
\begin{eqnarray}
\label{eq52}
A \!\!&=&\!\! -8\pi G\, (2\sigma+p), \\
%\label{eq53}
%
B \!\!&=&\!\! 8\pi G\, (\sigma+p), \\
\label{eq54}
\alpha \!\!&=&\!\! 2\pi G\, \sigma, \\
\label{eq55}
c_{1} \!\!&=&\!\! \pi G\, (3\sigma + 4 p).
\end{eqnarray}
We note that the metric (\ref{eq39})-(\ref{eq41}) does not match
with the metric (\ref{eq7})-(\ref{eq9}) for any values of the
constants $A$, $B$, $\alpha$ and $\beta$.
Finally, in the weak-field approximation, the metrics are:
\begin{eqnarray}
\label{eq56}
ds^2  \!\!&=&\!\! [1 \,+ \, 8\pi G\, (2\sigma+p)\, t \, + \,
4\pi G\, (\sigma+2p)|x|\,]\,dt^2 \nonumber \\
 \!\!&-&\!\! [1 \,
+ \, 8\pi G\, (\sigma+p)
\,t \, - \, 4\pi G\, (\sigma-2p) |x|\,]\,dx^2 \nonumber \\
 \!\!&-&\!\! [1 \, + \, 4\pi G\, \sigma \,t \, -4\pi G\, \sigma|x|\,] \,
 (dy^2+dz^2),
\end{eqnarray}
in the second case and
\begin{eqnarray}
\label{eq57}
ds^2  \!\!&=&\!\! [1 \,- \, 8\pi G\, (2\sigma+p)\, t \, + \,
4\pi G\, (\sigma+2p)|x|\,]\,dt^2 \nonumber \\
 \!\!&-&\!\! [1 \,
- \, 8\pi G\, (\sigma+p)
\,t \, - \, 4\pi G\, (\sigma-2p)|x|\,]\,dx^2 \nonumber \\
 \!\!&-&\!\! [1 \, - \, 4\pi G\, \sigma \,t \, -4\pi G\, \sigma|x|\,] \,
(dy^2+dz^2),
\end{eqnarray}
in the third case. We observe that it is possible to obtain the
metric (\ref{eq57}) by time-reversing the metric (\ref{eq56}).
\\
It is interesting to calculate acceleration of a test particle
initially at rest at small distance from wall plane.
\\
The geodesic motion  of the particle is given by
\begin{equation}
\label{eq58}
U^{\alpha} \, {U^{\beta}}_{; \; \alpha}=0.
\end{equation}
where $U^{\alpha}$ is the four velocity.
We wish to find the acceleration $a^{\alpha}= d U^{\alpha} / d
\tau$; moreover we assume the particle initially at rest, so that
$U=(1,0,0,0)$. The geodesic equation, under these initial
conditions becomes $d U^{\alpha} / d \tau = - {\Gamma^{\beta}}_{0
\; 0}$. Therefore the acceleration is
\begin{equation}
\label{eq59}
a^x= - \frac {2 \; \pi \;  G \; (\sigma+2p)\; \text{sign}\, x}
{1+8 \pi G (\sigma+p) t + 4 \pi G (\sigma - 2 p)} \, .
\end{equation}
that in weak field approximation and in the case $x>0$, becomes
\begin{equation}
\label{eq59bis}
a^x= - 2 \; \pi \;  G \; (\sigma+2p).
\end{equation}
Let us analyze this equation in three different cases.
\\
\\
1) {\it Domain walls}.
\\
Domain walls are two dimensional surfaces interpolating between
separate vacua with different vacuum expectation value of the
scalar field. They are produced through the breakdown of discrete
symmetry. These defects are formed by the Kibble mechanism
\cite{Kibble} when different regions cool in a hot Universe into
different minima of the potential. They have been studied
intensively in the past, mainly because of their notable
cosmological implications \cite{VILENK}. Domain walls are
characterized by an equation of state for which the surface energy
density is equal to minus the surface pressure
\begin{equation}
\label{eq60} \sigma = -p \, .
\end{equation}
Therefore the acceleration of the test particle is
\begin{equation}
\label{eq61} a^x = 2 \pi G \sigma \;
> 0
\end{equation}
that is to say the domain walls repels a test particle placed near
the wall, the tension acts as a repulsive source of gravity
whereas the pressure  is attractive (see also \cite{gron}). The
cosmological consequences of domain walls is strictly connected to
these unusual and interesting source of "antigravity". In fact we
are faced with the recent problem of understanding a Universe with
an equation of state $w < -1/3$ in order to explain the nature of
the so called dark energy of the Universe. From this point of
view, domain walls are natural candidates to obtain $w < -1/3$.
\\
\\
2) {\it Ferromagnetic domain walls}.
\\
These particular walls have been recently proposed to account the
primordial magnetic field in the Universe \cite{cea}. The
ferromagnetic domain walls are domain walls in which a uniform
magnetic condensate is localized on the wall, originating from a
peculiar magnetic condensation induced by fermion zero mode
localized on the wall. In a recent paper it has been investigate
the gravitational properties of these objects \cite{FOGLI}, and it
has been found that the equation of state is
\begin{equation}
\label{eq62}
p= \frac {\sigma} {2}.
\end{equation}
As a consequence  the acceleration is, from Eq.~(63)
\begin{equation}
\label{eq63}
a^x=- 4 \pi G \sigma <  0.
\end{equation}
The acceleration is negative, the particle accelerates toward the
wall.
\\
\\
3) {\it Shell dust}
\\
In this case we consider the plane formed by shell dust. The
pressure is null $p=0$, therefore the expression of the
acceleration is
\begin{equation}
\label{eq64} a^x= - 2 \pi G \sigma   < 0
\end{equation}
the plane is attractive as i the previous case.
\\
Topological defects could provide the seeds for the gravitational
clustering of galaxies and other structures. From this point of
view the most studied and promising defects are cosmic strings. In
fact cosmic strings carry energy density, curve space-time and can
act as the seed for the gravitational accretion, due to the
Newtonian gravitational attractive force on the surrounding
matter.
\\
In this context the domain wall do not give rise to a structure
formation model and are ruled out for two reasons: 1) stable
domain walls are a cosmological disaster, a single wall stretching
across the Universe would overclose it  \cite{ZELDOVICH}; 2)
domain walls act, as we already seen, in order to repel  matter
therefore they contribute to break up the seeds.
\\
On the other hand,  ferromagnetic domain walls are able to
overcome these two problems. The first problem regarding the so
called "domain wall problem" is solved because of the surface
energy density of these walls scales as $R^{-3}$ \cite {FOGLI}
therefore these defects do not give rise to  problems from a
cosmological point of view. The second problem is solved because,
as discussed  before, ferromagnetic domain walls are attractive
allowing the growth of cosmological perturbations in an expanding
Universe.
%
%###########################################################################
%
\vskip 2truecm
\section{Non-reflection-symmetric gravitational field of
ferromagnetic domain walls}
\vskip 1truecm
In  previous section we have studied the solutions, in the weak
field approximation, of the Einstein's equations for an
energy-momentum tensor given by Eq.(\ref{eq9}). As we have seen,
Vilenkin derived reflection-symmetric static  solutions for vacuum
domain walls and these solutions are inconsistent with the Kasner
solution\footnote{We remember that the Kasner metric is the most
general static solution of Einstein's equations in the vacuum
depending from one spatial coordinate.}. Tomita \cite{TOMITA}
studied solutions that are not-reflection-symmetric and he found
that these solutions are consistent with Kasner solution. In this
section we consider the static non-reflection-symmetric solutions,
in the general case of energy-momentum tensor given by
Eq.(\ref{eq9}). The solutions of Eqs.(\ref{eq5}),(\ref{eq6}) with
the tensor (\ref{eq9}) are
\begin{eqnarray}
\label{eq65}
 h_{00}(x) & = & 4 \pi G |x| \, (\sigma + 2 p) +
(\beta+\gamma-\alpha) x+ a_0 \nonumber \\
h_{11}(x) & = & 4 \pi G |x| \, (\sigma - 2 p) + \alpha \, x + a_1
\nonumber \\
h_{22}(x) & = & 4 \pi G |x| \, \sigma + \beta x + a_2
\\
h_{33} (x) & = & 4 \pi G |x| \,  \sigma + \gamma x + a_3 \nonumber
\end{eqnarray}
with $\alpha,\beta,\gamma, a_0,a_1,a_2,a_3$ constants that were
not present in the symmetric  Vilenkin's solutions. Now we compare
the solutions (\ref{eq65}) with the Kasner metric in the region
$x>0$, given by
\begin{equation}
\label{eq66} dS^2=X^{2 p_1} \, dT^2 - d X^2 - X^{2 p_2} \,  dY^2 -
X^{2 p_3} \, dZ^2,
\end{equation}
with
\begin{equation}
\label{eq67} p_1+p_2+p_3=1
\end{equation}
and
\begin{equation}
\label{eq68} p_1^2+p_2^2+p_3^2=1.
\end{equation}
Let us consider the following transformation of coordinate:
\begin{eqnarray}
\label{eq69} T & = & K^{p_1} \, t \nonumber
\\
X & = & K^{-1} \, e^{-Kx} \nonumber
\\
Y & = & K^{p_2} \, y
\\
Z & = &  K^{p_3} \, z \nonumber
\end{eqnarray}
with $K>0$, constant. Let us consider the region of small
gravitational potentials, namely the region where $0 < x \ll 1$.
The metric (\ref{eq66}), with the transformations (\ref{eq69}),
gives the following relations among harmonic coordinate:
\begin{eqnarray}
\label{eq70}
h_{00} & = & -2 \, p_1 \, K \, x \nonumber \\
h_{11} & = & 2 \,  K \,  x \nonumber \\
h_{22}  & = & 2 \, p_2 \, K \, x \\
h_{33} & = & 2 \, p_3 \, K \, x. \nonumber
\end{eqnarray}
If we comparing the metrics (\ref{eq65}) and (\ref{eq70}), we have
\begin{eqnarray}
\label{eq71}
& a_0 & = a_1=a_2=a_3=0 \nonumber \\
& K & = 2 \pi G (\sigma-2 p+\alpha) \nonumber \\
& p_1 & = - 2 \pi G K^{-1} ( \sigma + 2 p + \beta + \gamma - \alpha) \\
& p_2 & = 2 \pi G K^{-1} ( \sigma + \beta) \nonumber \\
& p_3 & = 2 \pi G K^{-1} (\sigma + \gamma). \nonumber
\end{eqnarray}
Moreover the conditions (\ref{eq67}) and (\ref{eq68}) give
\begin{eqnarray}
\label{eq72}
\sum_{i=1}^3 p_i & = & 1  \, \Longrightarrow \, \, \, \, \,
\sigma-2p+\alpha= \frac {K} {2 \pi G}
\\
\label{eq73}
\sum_{i=1}^3 p^2_i & = & 1 \, \Longrightarrow \, \, \,
(\sigma + 2 p + \beta + \gamma - \alpha)^2 + (\sigma+ \beta)^2 +
(\sigma + \gamma)^2 = \left( \frac {K} { 2 \pi G} \right)^2.
\end{eqnarray}
When we consider the case $x<0$, it is necessary to introduce
another Kasner metrics:
\begin{equation}
\label{eq74} ds^2={\bar {X}}^{2 {\bar{p_1}}} \,  d {\bar{T}}^2 - d
{\bar{X}}^{2} - {\bar{X}}^{2 {\bar{p_2}}} \, d {\bar{Y}}^2 -
{\bar{X}}^{2 {\bar{p_3}}} \, d {\bar{Z}}^2
\end{equation}
with
\begin{eqnarray}
\label{eq75}
{\bar{p}}_1 + {\bar{p}}_2 + {\bar{p}}_3  & = & 1
\\
\label{eq76}
{\bar{p}}_1^2 + {\bar{p}}_2^2 + {\bar{p}}_3^2  & = & 1
\end{eqnarray}
and the analogous coordinate transformations in Eqs. (\ref{eq69})
\begin{eqnarray}
\label{eq77}
\bar{T} & = & {\bar{K}}^{\bar{p}_1} \, t \nonumber
\\
\bar{X} & = & {\bar{K}}^{-1} \, e^{- \bar{K} x} \nonumber
\\
\bar{Y} & = & {\bar{K}}^{\bar{p}_2} \, y
\\
\bar{Z}  & = &  {\bar{K}}^{\bar{p}_3} \, z. \nonumber
\end{eqnarray}
As in the previous case,  considering small gravitational regions,
 $0< - x \ll 1$, we have
\begin{eqnarray}
\label{eq78}
h_{00} & = & 2 \, {\bar{p}}_1 \, K \, x \nonumber \\
h_{11} & = & - 2 \,  K \,  x \nonumber \\
h_{22}  & = & - 2 \, {\bar{p}}_2 \, K \, x \\
h_{33} & = & - 2 \, {\bar{p}}_3 \, K \, x. \nonumber
\end{eqnarray}
Analogously to the case  $x>0$, if we compare the metrics
(\ref{eq65}) and (\ref{eq78}) we have
\begin{eqnarray}
\label{eq79}
& a_0 & = a_1=a_2=a_3=0 \nonumber \\
& K & = 2 \pi G (\sigma-2 p-\alpha) \nonumber \\
& {\bar{p}}_1 & = - 2 \pi G K^{-1} ( -\sigma - 2 p - \beta + \gamma - \alpha) \\
& {\bar{p}}_2 & = 2 \pi G K^{-1} ( \sigma - \beta) \nonumber \\
& {\bar{p}}_3 & = 2 \pi G K^{-1} (\sigma - \gamma) \nonumber
\end{eqnarray}
and the conditions (\ref{eq75}) and (\ref{eq76})  give
respectively
\begin{eqnarray}
\label{eq80}
\sum_{i=1}^3 {\bar{p}}_i & = & 1  \, \Longrightarrow \, \, \, \,
\, \sigma-2p-\alpha= \frac {K} {2 \pi G}
\\
\label{eq81}
\sum_{i=1}^3 {\bar{p}}^2_i & = & 1 \, \Longrightarrow \, \, \,
(\sigma + 2 p - \beta - \gamma + \alpha)^2 + (\sigma - \beta)^2 +
(\sigma - \gamma)^2 = \left( \frac {K} { 2 \pi G} \right)^2.
\end{eqnarray}
Now, comparing  the second equation in  Eq.(\ref{eq71}) with the
second equation in  Eq.(\ref{eq79}), and the second equation in
Eq.(\ref{eq74}) with the second equation in Eq.(\ref{eq81}) we
get:
\begin{eqnarray}
\label{eq82}
\alpha & = & 0  \\
\label{eq83}
2 \pi G (\sigma - 2 p) & = &  K \\
\label{eq84}
(\beta + \gamma) (\sigma + p) & = & 0.
\end{eqnarray}
Therefore, if Eqs.(\ref{eq82})-(88) are satisfied, then it is
possible to identify the metrics in the regions $0< x \ll 1$ and
$0 < -x \ll 1$. For completeness we write the relations between
$p_i$ and ${\bar{p}}_i$:
\begin{eqnarray}
p_1+{\bar{p}}_1 & = & - \frac {2 (\sigma+ 2 p)} {\sigma - 2
p}, \\
p_2+{\bar{p}}_2 =   p_3+{\bar{p}}_3 & = &  \frac {2 \sigma}
{\sigma - 2p}.
\end{eqnarray}
Note that  Tomita's results are recovered  if we consider the case
of a vacuum domain wall: $\sigma+p=0$. If we impose spatial
reflection symmetry with respect to the plane of the wall, we
obtain the Vilenkin's case, $4p+\sigma=0$. But it is important to
observe that from Eq.(\ref{eq83}), in order to have static
solution, it must be $K>0$, namely   $p< \sigma /2$. In the case
of ferromagnetic domain walls, we have $p= \sigma /2$. So that
there are not non-reflection static solutions for ferromagnetic
domain walls.
%*************************************************************************
%*************************************************************************
\section{conclusion}
\vskip 1truecm
In this paper we have discussed the gravitational properties  of
plane walls in the weak field approximations. In this
approximation we found the general time-dependent metric. We have
discussed the case of domain walls where an observer experience a
repulsion from the wall. On the other hand, ferromagnetic domain
walls and shell dust are gravitationally attractive. In other
words, the plane exercises  a gravitational attraction that could
make object interesting for the formation of large scale structure
in the Universe.
We have obtained the non-static metrics
Eqs.~(\ref{eq56}),(\ref{eq57}), in the weak-field approximation
from the exact metric of an homogeneous massive plane \cite{TAUB}.
In particular we have obtained for the coefficients $A$ and $B$,
in Eq. (\ref{eq10}), the following values: $A= \pm 8 \pi G (2
\sigma +p)$ and $B= \mp 8 \pi G (\sigma +p)$. Therefore to get a
static solution  it is necessary that $A=B=0$, and this is
obtained only when $\sigma=p=0$ which  is obviously not physically
relevant. Therefore we conclude that it is impossible to obtain
static metric in weak-field approximation \cite{VILENKIN:81} in
the case of plane vacuum domain walls ($\sigma=-p\neq 0$)
according to \cite{VILENKIN:83}. Moreover we have found that there
is not static non-reflection-symmetric gravitational field for
ferromagnetic domain walls, while is possible in the case of
domain walls as already pointed out in \cite {TOMITA} have been
already studied by Tomita.
%
%###############################################################################
%****************************************************************

%
%
%
%
%
%
\end{document}